# Calculation of binding energies and radii of proton-rich nuclei

I. Casinos

*Facultad de Química, Universidad del País Vasco, Pº Manuel de Lardizabal 3, 20018 San Sebastián, Spain*

e-mail: ismael.casinos@ehu.es

## Abstract

The n,p-networks model is used to estimate binding energies and radii of proton-rich ($N \leq Z$) nuclei. These calculations have been made for some representative examples of even-$Z$ and odd-$Z$ nuclei with nucleon numbers lower than sixty. Good agreement is found in the comparison of these results with experimental data.

## 1 Introduction

A lot of effort has been made in the development of theoretical models able to reproduce and predict the properties and behaviour of nuclei throughout the periodic table.

Calculation of the nuclear binding energy is useful as a test to assess the goodness of a model approach to study the atomic nucleus [1-5]. In a previous report [6], I have suggested a nuclear potential and a model, in a non-quantum mechanics approach, to study the binding energy and radius of $N=Z-1$, $Z$, $Z+1$ nuclei with nucleon numbers lower than sixty. And it was taken as a base to study the fusion reaction of two nuclei. This model was extended [7] in order to calculate these nuclear properties for some exemplary neutron-rich nuclei with mass numbers lower than sixty.

The study of proton-rich nuclei has attracted the attention of researchers [8-20] in order to reach an understanding of the peripheral structure of these nuclei.

The aim of this report is to make use of the cited model to estimate binding energies and radii of several $N \leq Z$ nuclei.

## 2 Short sketch of the model

In this section, a partial account of the model is accomplished. A detailed picture can be found in Ref. [6].

In order to estimate the binding energy of a nucleus C, $E_b(C)$, the whole formation process from its nucleons $Zp + Nn \rightarrow C(Z, N)$ is considered by means of the successive and appropriate nucleon-additions $A+n \rightarrow C$ and $A+p \rightarrow C$ following Scheme 1 with the associated addition (separation) energies $S_n$ and $S_p$, respectively. In other words, a nucleus C formed in a particular nucleon addition

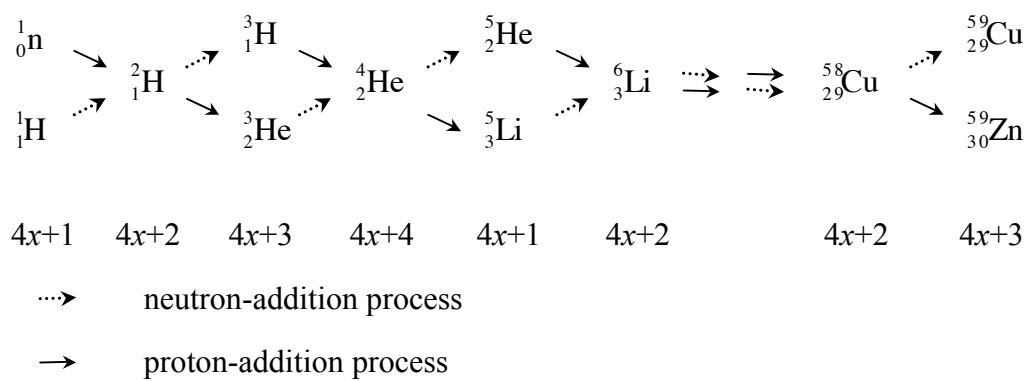

$4x+1 \quad 4x+2 \quad 4x+3 \quad 4x+4 \quad 4x+1 \quad 4x+2 \quad\quad\quad 4x+2 \quad 4x+3$

⋯▸     neutron-addition process

→     proton-addition process

$x = 0, 1, 2, \ldots 14$.

to a nucleus A is held as nucleus A taking part in the next addition.

$$S_n = -\frac{kM_A(R_A + R_n)^6}{8R_A^3 R_n^3 D_C^6} \tag{1}$$

$$S_p = -\frac{kM_A(R_A + R_p)^6}{8R_A^3 R_p^3 D_C^6} + \frac{K_e e^2 Z_A}{D_C} \tag{2}$$

$M_A$ and $Z_A$ are nucleon and proton numbers of nucleus A, respectively. $R_n = R_p =$ 1.1498 fm are neutron and proton radii. The diffuse surface radius of nucleus A, $R_A = R_{0A} M_A^{1/3}$, is the distance where its nucleonic distribution starts to be affected by the interaction with an approaching nucleon (or nucleus). The kernel radius of nucleus C, $D_C = aM_A^{1/3} + b$ (where $M_A = M_C - 1$), denotes the location distance, related to an energy minimum, of the outermost nucleon constituting the globular nucleus in its ground state. Values for all these parameters are displayed in Table 1 to be appropriately applied in eqs. (1) and (2) for each successive nucleon addition along Scheme 1 to perform a nucleus C formation.

**Table 1.** Parameter values to be used in eqs. (1) and (2).

| Type of nuclide A | Type of nuclide C | $a$ (fm) | $b$ (fm) | $R_{0A}$ (fm) | $k$ (MeV fm$^6$) |
|---|---|---|---|---|---|
| 4x+4 | 4x+1 | 0.68994 | 1.5116 | 2.9271 | 4.8798 |
| 4x+1 | 4x+2 | 0.71841 | 1.3948 | 1.1858 | 41.129 |
| 4x+2 | 4x+3 | 0.87530 | 0.82888 | 2.2436 | 11.778 |
| 4x+3 | 4x+4 | 0.92065 | 0.64997 | 1.2772 | 43.113 |

## 3 Calculation and results

This model regards a nucleus as formed by two interpenetrated similar networks of neutrons and protons displaying a dependence of the interaction strength, $k$, (see Table 1) on the *np, nn, pp* pairings between nucleons that favours the formation of compact α-particle clusters jointly to residual valence nucleons.

The binding energy and radius of $N = Z$-1, $Z$, $Z$+1 nuclei with nucleon numbers lower than 60 displayed in the preceding Scheme 1 were calculated in this model [6]. Moreover, these nuclear properties were also calculated for some representative $N \geq Z$ nuclei by considering a high-density core of nucleons constituted by α-particle clusters and a low-density skin of neutrons where each neutron makes use of one α-particle space [7]. Complementarily, in this work it is intended to perform those calculations for some exemplary $N \leq Z$ nuclei by assuming the same nucleonic distribution for mirror nuclei (i.e., $Z$ and $N$ values for a nucleus are the same that the $N$ and $Z$ values for its mirror, respectively).

The assumed nucleonic distribution in $N \leq Z$ nuclei is described below by taking two illustrative examples: the silicon isotopes as even-$Z$ nucleus and the phosphorus isotopes as odd-$Z$ nucleus.

$^{24}_{14}$Si is built by the alternate addition of ten neutrons and ten protons following Scheme 1 forming five α-particle clusters that compose the core of $^{24}_{14}$Si jointly to its skin which is made by the successive addition with the lowest interaction strength ($k$ = 4.8798 MeV fm$^6$) of the four residual protons that make use of four consecutive α-particle spaces.

$^{25}_{14}$Si is composed by five α-particle clusters jointly to the incomplete 6$^{th}$ one (made of 11$^{th}$ proton, 11$^{th}$ neutron and 12$^{th}$ proton) and both 7$^{th}$ and 8$^{th}$ α-particle spaces with 13$^{th}$ and 14$^{th}$ protons, respectively.

$^{26}_{14}$Si is composed by six α-particle clusters jointly to 13$^{th}$ and 14$^{th}$ protons that make use of two consecutive α-particle spaces.

$^{27}_{14}$Si is composed by six α-particle clusters jointly to the incomplete 7$^{th}$ one made of 13$^{th}$ proton, 13$^{th}$ neutron and 14$^{th}$ proton.

$^{28}_{14}$Si is composed by seven α-particle clusters.

$^{26}_{15}$P is built by the alternate addition of eleven neutrons and twelve protons following Scheme 1 forming five α-particle clusters and the incomplete 6$^{th}$ one (made of 11$^{th}$ proton, 11$^{th}$ neutron and 12$^{th}$ proton) that compose the core of $^{26}_{15}$P jointly to its skin which is made by the successive addition with the lowest interaction strength ($k$ = 4.8798 MeV fm$^6$) of the three residual protons that make use of three consecutive α-particle spaces.

$^{27}_{15}$P is composed by six α-particle clusters jointly to 13$^{th}$, 14$^{th}$ and 15$^{th}$ protons that make use of three consecutive α-particle spaces.

${}^{28}_{15}$P is composed by six α-particle clusters jointly to the incomplete 7$^{th}$ one (made of 13$^{th}$ proton, 13$^{th}$ neutron and 14$^{th}$ proton) and the 8$^{th}$ α-particle space with the 15$^{th}$ proton.

${}^{29}_{15}$P is composed by seven α-particle clusters and the 8$^{th}$ α-particle space with the 15$^{th}$ proton.

${}^{30}_{15}$P is composed by seven α-particle clusters and the 8$^{th}$ α-particle space with both 15$^{th}$ proton and 15$^{th}$ neutron.

Calculations of binding energy per nucleon are shown in Fig. 1 and those of kernel radius in Fig. 2. And these results are displayed in Fig. 1(a) and Fig. 2(a) for even-$Z$ nuclei (isotopes of carbon, silicon and titanium), and in Fig. 1(b) and Fig. 2(b) for odd-$Z$ nuclei (isotopes of nitrogen, phosphorus and vanadium).

As can be seen in Fig. 1, the calculated binding energies for nuclei with a high ratio of nucleons in their high-density core to protons in their low-density skin (i.e., $N \approx Z$) are better adapted to experimental data [21] than those with a low ratio (i.e., $N < Z$). Consequently, a variation in a uniform nuclear density, mainly in the second case, could affect the spherical radius and the interaction strength between two nuclear species along a nucleus formation that have not been taken into account in these calculations.

In conclusion, it is appreciated a concurrence between calculated and experimental data that sustains the usefulness and development of this model in nuclear studies.

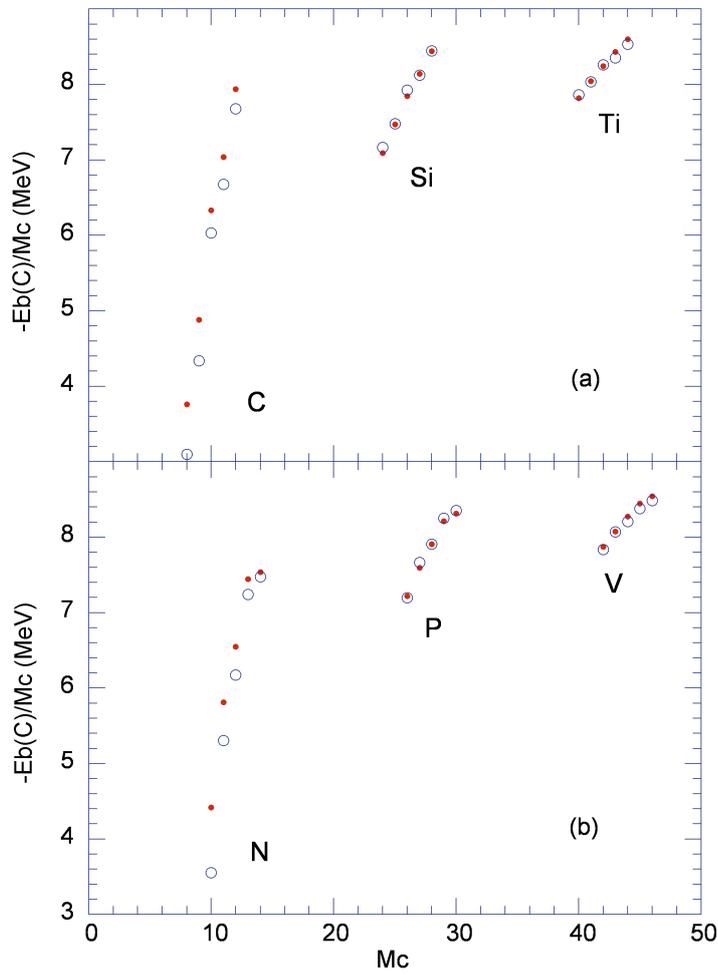

**Fig.1.** Calculated (filled circles) and experimental (unfilled circles) binding energies per nucleon for $N \leq Z$ nuclei. (a), even-$Z$ nuclei. (b), odd-$Z$ nuclei.

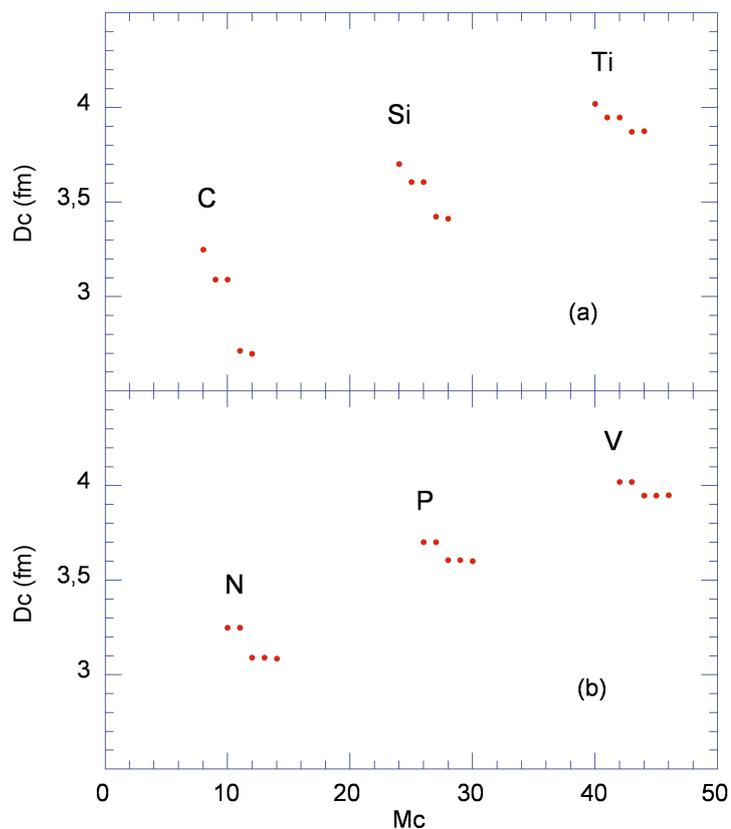

**Fig.2.** Calculated kernel radii for $N \leq Z$ nuclei. (a), even-$Z$ nuclei. (b), odd-$Z$ nuclei.